\documentstyle[aps,prd,epsfig]{revtex}
\tighten

\newcommand{\mb}[1]{\mbox{\boldmath $#1$}}

\begin{document} 
 
\title{Evolution systems for non-linear perturbations of background
geometries}

\author{Philippos Papadopoulos and Carlos F. Sopuerta}

\address{~}

\address{Relativity and Cosmology Group, School of Computer Science
and Mathematics, \\
Portsmouth University, Portsmouth~PO1~2EG, Britain}

\address{~}

\date{\today}

\maketitle 
 
\begin{abstract}
The formulation of the initial value problem for the Einstein
equations is at the heart of obtaining interesting new solutions using
numerical relativity and still very much under theoretical and applied
scrutiny. We develop a specialised background geometry approach, for
systems where there is non-trivial a priori knowledge about the
spacetime under study.  The background three-geometry and associated
connection are used to express the ADM evolution equations in terms of
physical non-linear deviations from that background. Expressing the
equations in first order form leads naturally to a system closely
linked to the Einstein-Christoffel system, introduced by Anderson and
York, and sharing its hyperbolicity properties.  We illustrate the
drastic alteration of the source structure of the equations,
and discuss why this is likely to be numerically advantageous.
\end{abstract}
  
 

\section{Introduction}\label{sec1}

An ongoing research effort, numerical relativity, attempts to solve
the partial differential equations (PDEs) pertaining to the Einstein
equations using numerical methods. A significant current focus is
centered on the study of the relativistic two-body
problem~\cite{BBH}. The broad features of the dynamics are already
{\em approximately} known, through well established semi-analytical
approaches, including perturbation theory~(see e.g., a review in
\cite{PULLIN}) and post-Newtonian approximations~(see e.g., latest
developments in \cite{DAMOUR} and references therein).  It seems
intuitively right that one should be able to leverage such knowledge
and reduce the computational complexity of the problem. There have
been recently a number of proposals on various non-standard {\em
physical approximations} to the spacetimes involved in
mergers~\cite{WILSON,THORNE,LAGUNA,LAZARUS}. A common theme among such
proposals is the elimination of degrees of freedom which are thought
of as not crucial for some phase of the computation.  There is also
presently activity on the mathematical notions underlying numerical
relativity, with particular emphasis on the hyperbolicity properties
of the equations (see e.g., \cite{FRIE}) and the nature of the
propagation of the constraints.  Even while the mathematical structure
of the general problem is under scrutiny, there are particular
instances of initial value problems which provide significant
insights. For example, characteristic evolution of spacetimes that
admit caustic-free foliations appear to be completely
tractable~\cite{BBH2}. Significant improvements have also been
claimed using approaches which decompose the field variables in
certain ways~\cite{NAKA,SHAP}.

We would like to advance in this paper a different perspective for
numerical relativity. We will avoid any physical approximation to the
equations, but we will try instead to pin down geometrically and
analytically the system by introducing additional ``guiding''
structures in the spacetime manifold. The key concept will be a
background geometry, constructed analytically, or semi-analytically,
and encompassing the a priori knowledge about the system. The true
dynamics will then unfold as a ``correction'' on the background. The
proposal is promising on several grounds. It is easy to show that
reformulating the evolution in terms of finite perturbations can lead
to significant gains in {\em accuracy} for systems that never deviate
significantly from the background. This is also borne out in
preliminary computations. It is possible that the {\em stability} of
computations will improve as well. This is much harder to ascertain
concretely. We note that currently the main avenues for improving
stability are sought in the direction of securing hyperbolicity
properties and the convergence of the constraints, but this effort has
not been conclusive. The relevance of source terms seems less well
understood.

The concept of a background geometry has already been used in analytic
studies of initial value problems (see e.g., in~\cite{HAWE}), and it has
also been claimed that the use of a background spacetime clarifies the
task of defining energy and momentum quantities for the gravitational
field~\cite{GRIS}.  The treatment presented in~\cite{HAWE} is
particularly enlightening for the motivation of this paper.  There, 
two four-dimensional metrics are introduced on the
spacetime manifold $M$. One of them, say $g_{\mu\nu}$, is promoted
to physical status by virtue of satisfying the Einstein equations. The
other, $\bar{g}_{\mu\nu}$, is relegated to a background role. Its
associated Christoffel symbols, $\bar{\Gamma}^{\mu}{}_{\nu\rho}$, are
used for the construction of a covariant derivative $\bar{\nabla}_{\mu}$. 
It is this derivative with respect to which one then expresses all 
relevant differential equations, e.g., gauge conditions and the 
Einstein equations. The fundamental evolution variables are the 
{\em inverse metric differences} 
$\chi^{\mu\nu}= g(g^{\mu\nu}-\bar{g}^{\mu\nu})\,,$ 
where $g$ is the determinant of the $g^{\mu\nu}$ metric. With the
imposition of the harmonic gauge, $\bar{\nabla}_{\mu}\chi^{\mu\nu} = 0\,,$
one obtains the reduced set of Einstein's equations,
\[ g^{\rho\sigma} \bar{\nabla}_\rho \bar{\nabla}_\sigma \chi^{\mu\nu} +
\mbox{L.O.T.}^{\mu\nu} = - 2 \bar{R}^{\mu\nu} + \bar{g}^{\mu\nu}\bar{R}\,,\]
where $\mbox{L.O.T.}^{\mu\nu}$ denotes terms of order lower than two
and barred quantities in the right-hand side are the Ricci curvatures
and scalars associated with the four dimensional background
metric. Two remarks are worthwhile here. First, if $\bar{g}_{\mu\nu}$
is {\em any} exact solution of $\bar{R}^{\mu\nu}=0$ then the RHS of
this equation is zero, which reflects the importance of the background
for the structure of the source terms.   Second, the hyperbolic 
characteristic structure of the equation is transparent, given its 
resemblance to a system of wave equations.  It is noteworthy that 
whereas the source structure is background dependent, the hyperbolic 
structure solely relies on the {\em physical} metric.

Following this example, we will formulate here the Einstein evolution 
problem with a view towards numerical applications. To this end we
will work within the framework of the 3+1 decomposition. There are
several reasons for opting for this path in view of the above
mentioned (covariant) examples. Maybe most importantly, we would like
to have explicit control over the gauge functions, as we would like to
construct them using a priori information. Additionally, the 3+1
decomposition is tightly linked to a Hamiltonian framework. We will
not explicitly exploit this feature here, but this will ultimately
facilitate a systematic use of background spacetimes derived from
post-Newtonian approximate solutions.

The main result of this paper is the introduction of a set of new 3+1
evolution systems for the Einstein equations which explicitly separate
a background geometry. The sets of evolution variables denote the
finite deviations of the physical geometry from the assumed
background. We derive systems involving second or first order spatial
derivatives, with a corresponding change in the number of
variables. In both cases we write the equations in a form which can
explicitly cancel stationary or quasi-stationary balance terms, {\em
provided that adequate a priori description of those terms exists}. We
found that implementing the background geometry concept in conjunction
with a hyperbolic system requires a slightly generalised geometrical
construction. We will hence talk about the strong and weak versions of
the proposal, according to the strength of the assumptions on the
specifications of the physical and background spacetimes.  All systems
are constructed as three-covariant with respect to the background metric.

The structure of the paper is as follows. In section II we review some
results in order to establish the notation. In Section III we will
describe a framework based on the introduction of a background spatial 
metric and develop the geometric objects emerging in this picture. We then 
write evolution equations for the spatial tensors describing the deviation 
of the physical geometry from a background one. The two natural sets 
emerging are closely related to the ADM and Einstein-Christoffel systems. 
We argue that the metric compatibility condition in the latter is more
properly seen as a geometric identity on the three dimensional Riemann
tensor. In Section IV we will generalise the framework slightly, in
order to develop a symmetric hyperbolic system for appropriate
evolution fields. In Section V we make a number of remarks.  We discuss
implementation issues and a number of potential applications.

\section{Preliminaries and Notation}

We will adopt the signature conventions used in~\cite{MTW}. Throughout 
the paper, Greek indices denote spacetime indices running from 0 to 3, 
whereas Latin indices are spatial indices and run from 1 to 3. Parentheses 
and square brackets will denote symmetrization and antisymmetrization
respectively. Reference to geometric objects like vectors and tensors
will refer to spatial objects, unless otherwise stated. We will
systematically use an overbar to distinguish geometrical objects
referring to a background geometry. Raising and lowering of indices
will always be using the physical metric. We will consider only the
vacuum case, i.e., absence of matter sources ($T_{\mu\nu}=0$). Studies
of matter systems (in particular fluids) along similar lines are
pursued separately~\cite{SPA} and a generalisation will be given 
elsewhere.

The 3+1 decomposition (also known as the Arnowitt-Deser-Misner (ADM)
formulation) of Einstein's equations~\cite{ADMF} has been discussed in
detail in many works (see, e.g.,~\cite{ADMD}). It assumes that the
spacetime has topology $R \times \Sigma$, and hence it is foliated by
the level surfaces $\Sigma(t)$ of a scalar function $t(x^\mu)$.  The
unit normal to the hypersurfaces $\Sigma(t)$, $n^\mu=-N\nabla^\mu t$,
satisfies $g_{\mu\nu}n^{\mu}n^{\nu}+1=0$. It can be used to decompose the 
four-metric as
\begin{eqnarray}
  g_{\mu\nu} = h_{\mu\nu}  - n_{\mu}n_{\nu} \,. \nonumber
\end{eqnarray}
The vector field $t^{\mu}$ (time threading of the manifold,
$t^{\mu}\nabla_{\mu}t=1$) giving rise to the relations $N = - t^{\mu}
n_{\mu}$ and $\beta^{\mu} = t^{\mu} - N n^{\mu}$ allows to write the
metric element in the form
\begin{eqnarray}
ds^2 = -N^2dt^2+h_{ij}(dx^i+\beta^idt)(dx^j+\beta^jdt) \,. \nonumber
\end{eqnarray}
Here, $N$ is the lapse scalar, which is assumed to be positive, 
$\beta^i$ is the spatial shift vector,
and $h_{ij}$ is the spatial metric (or first fundamental form) induced
on the hypersurfaces $\Sigma(t)$.

In terms of the unit normal, the second fundamental form or extrinsic 
curvature of the spatial hypersurfaces is given by
\begin{equation}
K_{ij}= -\textstyle{1\over2}{\pounds}_n h_{ij}~~
\Rightarrow ~~ \hat{\partial}_t h_{ij} = -2NK_{ij} \,, \label{hdot}
\end{equation}
where the time operator is $\hat{\partial}_t =
\partial_t-{\pounds}_\beta$, and ${\pounds}_\beta$ denotes Lie
differentiation with respect to $\beta^i$.  In particular,
\begin{eqnarray}
 {\pounds}_\beta h_{ij}  = \beta^k\partial_k h_{ij} +
h_{ik}\partial_j\beta^k+h_{jk}\partial_i\beta^k  = 2 D_{(i} \beta_{j)} \,,
\nonumber
\end{eqnarray}
where $\beta_i = h_{ij}\beta^j$. 
The Einstein equations are decomposed in a set of evolution equations
for the components of the extrinsic curvature
\begin{equation}
\hat{\partial}_t K_{ij} = -D_iD_j N + N( R_{ij}+KK_{ij}-
2K_i{}^lK_{jl}) \,, \label{kdot}
\end{equation}
the momentum constraint
\begin{equation}
D_j K^j{}_i - D_i K = 0 \,, \label{mcon}
\end{equation}
and the Hamiltonian constraint
\begin{equation}
R + K^2 - K^{ij}K_{ij} = 0 \,. \label{hcon}
\end{equation}
In these equations $K=h^{ij}K_{ij}$, $D_i$, $R_{ij}$, and $R$ are
the canonical covariant derivative, Ricci tensor, and scalar curvature
associated with the spatial metric $h_{ij}$ respectively.  The
Einstein field equations are equivalent to the evolution equations
(\ref{hdot}) and (\ref{kdot}) together with the constraints
(\ref{mcon}) and (\ref{hcon}). Given initial data on an initial
hypersurface $\Sigma(t_o)$, i.e.  $(h^o_{ij},K^o_{ij})$ satisfying the
constraints, and freely specifiable quantities $N$ and $\beta^i$
defined for all $\Sigma(t)$, one can use the evolution equations to
find the future development of such data. The framework is explicitly
three-covariant.

\section{3+1 decomposition with background metric: Strong version}\label{sec2}

We assume a second spatial metric field, denoted by $\bar{h}_{ij}$,
cohabiting with $h_{ij}$ on each of the hypersurfaces $\Sigma(t)$. The
four-dimensional manifold $M$ is then endowed with an additional
four-metric defined by
\begin{eqnarray}
\bar{g}^{\mu\nu}  =  \bar{h}^{\mu\nu} - n^{\mu}n^{\nu}\,, \nonumber
\end{eqnarray}
where
\begin{equation}
\bar{h}^{\mu\nu}n_\nu =0 \hspace{4mm} \mbox{and} \hspace{4mm} \bar{h}^{ij}= 
(\mb{\bar{h}^{-1}})^{ij}\,. \label{conhb}
\end{equation}
The normal $n^{\mu}=g^{\mu\nu}n_\nu$ is now a unit normal vector with respect 
to {\em both} metrics.  Then, there is a unique projection operator onto the 
hypersurfaces $\Sigma(t)$, $h^{\mu}{}_{\nu}=\bar{h}^{\mu}{}_{\nu}=
\delta^{\mu}{}_{\nu} + n^{\mu}n_{\nu}$.  As is clear, this construction uses
the same lapse $N$ and shift vector $\beta^i$ for both spacetimes and hence
the name {\em strong version}.
Taking into account the relations $N = -t^{\mu} n_{\mu}$ and 
$\beta^{\mu} = t^{\mu} - N n^{\mu}$ we write the background metric 
in the form
\begin{eqnarray}
ds^2 = -N^2dt^2+\bar{h}_{ij}(dx^i+\beta^idt)(dx^j+\beta^jdt) \,. \nonumber
\end{eqnarray}
The extrinsic curvature associated with the background geometry can be 
computed from the expression 
\begin{equation}
\bar{K}_{\mu\nu} = -\bar{h}_{(\mu}{}^{\rho}\bar{h}_{\nu)}{}^{\sigma}
\bar{\nabla}_\sigma n_\rho \,,  \label{caex}
\end{equation}
where $\bar{\nabla}_\mu$ is the connection associated with the background
spacetime metric $\bar{g}_{\mu\nu}$.  This leads to
\begin{eqnarray}
\bar{K}_{ij}= -\textstyle{1\over2}{\pounds}_n \bar{h}_{ij}~~
\Rightarrow ~~ \hat{\partial}_t \bar{h}_{ij} = -2N \bar{K}_{ij} \,. \nonumber
\end{eqnarray}

The two spatial metrics give rise to compatible covariant derivatives
$D_i$ and $\bar{D}_i$ satisfying $D_i h_{jk}=0$ and $\bar{D}_i 
\bar{h}_{jk} = 0$.  
If $\bar{\Gamma}^i{}_{jk}$ denotes the Christoffel symbols associated
with $\bar{h}_{ij}$ then $ {\cal E}^i{}_{jk} =
\Gamma^i{}_{jk}-\bar{\Gamma}^i{}_{jk} $, the difference of the
Christoffel symbols, is also a tensor.  Indeed, under a coordinate
change $x^i \rightarrow x^{i'} = f^{i'}(x^i)$, the Christoffel symbols
transform as
\[ \Gamma^{i'}{}_{j'k'} = \frac{\partial x^{i'}}{\partial x^i}
\Gamma^i{}_{jk}
\frac{\partial x^j}{\partial x^{j'}}\frac{\partial x^k}{\partial x^{k'}}+
\frac{\partial x^{i'}}{\partial x^l}\frac{\partial^2 x^l}
{\partial x^{j'}\partial x^{k'}} \,, \]
whence the tensorial transformation law for ${\cal E}^i{}_{jk}$ follows
\[ {\cal E}^{i'}{}_{j'k'}=
\frac{\partial x^{i'}}{\partial x^i}{\cal E}^i{}_{jk}
\frac{\partial x^j}{\partial x^{j'}}\frac{\partial x^k}{\partial x^{k'}}\,.\]
In fact, the explicitly covariant expression for
${\cal E}^i{}_{jk}$ is
\begin{equation}
{\cal E}^i{}_{jk} = \textstyle{1\over2}h^{il}(\Delta_{jkl}+
\Delta_{kjl}-\Delta_{ljk})  \,,
\label{eps}
\end{equation}
where
\begin{equation}
 \Delta_{kij} = \bar{D}_k h_{ij}\,.
\label{delt}
\end{equation}
We see that $\Delta_{kij}$ is the covariant derivative of the
physical spatial metric with respect to the background spatial
canonical connection.  The two covariant derivatives acting on a
general tensor field are related by
\begin{equation} 
D_i{\cal A}^{j\cdots}{}_{k\cdots} = 
\bar{D}_i{\cal A}^{j\cdots}{}_{k\cdots} 
+ {\cal E}^j{}_{il}{\cal A}^{l\cdots}{}_{k\cdots}+\cdots 
- {\cal E}^l{}_{ik}{\cal A}^{j\cdots}{}_{l\cdots}-\cdots \,. \label{dtdb}
\end{equation}

\begin{figure}
\centerline{\epsfig{figure=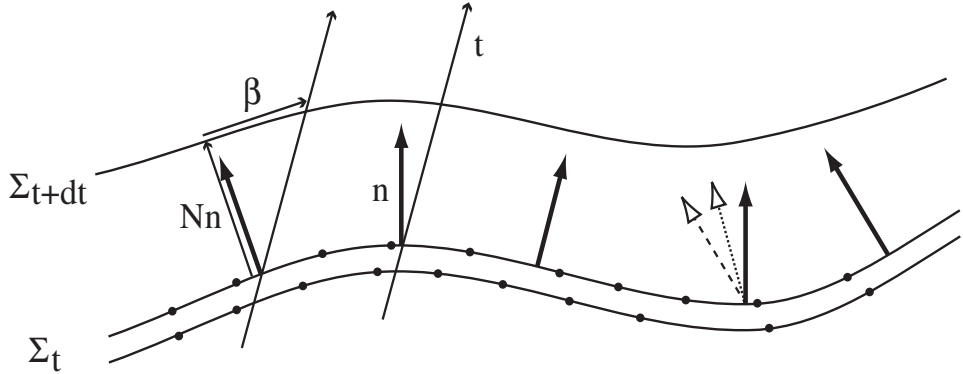,width=5in,height=2.0in}}
\vspace{0.5cm}
\caption{3+1 decomposition and a background geometry. Two dimensions
are suppressed for practical reasons. Two successive level surfaces of
the time scalar are shown, with corresponding unit normal vectors in
bold arrows.  The standard decomposition of the time flow vector field
into the lapse and shift components are shown. All those elements are
independent of the intrinsic 3-metric. The two different intrinsic
geometries are also shown, as two different copies of hypersurface
$\Sigma(t)$ (which should be thought as overlapping). The filled
circles on the hypersurface can be thought of as markers of equal
proper length, thus indicating the different intrinsic curvature.  The
effects on the {\em extrinsic curvature} are also indicated. The
right-most normal vector, when parallel transported to the left using
the physical metric produces the dashed vector.  The difference of
this vector from the local normal (bold vector) is the physical
extrinsic curvature.  The same original normal parallel transported
under the background metric, gives rise to the dotted vector, implying
a different (background) extrinsic curvature.}\label{fig1}
\end{figure}

Pictorially our geometric construct is presented in
Fig.~\ref{fig1}. The unit normal vector field $n^{\mu}$ represents
Eulerian observers momentarily at rest on the hypersurface $\Sigma(t)$
and is by construction the same for both spacetimes. 
The {\em acceleration} 1-form of those observers,
$a_{\mu}=n^{\nu} \nabla_{\nu}n_{\mu}$, (with $a^{\mu}n_{\mu}=0$), is
the same as seen from both spacetimes.  Some calculations show that
\begin{eqnarray}
\bar{a}_{\mu}= \bar{h}_{\mu}{}^{\nu}\bar{\nabla}_{\nu}(\ln N) =
h_{\mu}{}^{\nu}\nabla_{\nu}(\ln N) = a_{\mu} \,, \nonumber
\end{eqnarray}
where we have used the fact that the background and physical four-covariant
derivative have the same action on the scalar $\ln N$.

\subsection{Time evolution equations and constraints}

We turn now to the task of writing the 3+1 decomposed Einstein
equations in terms of a background metric. One further identity we
will need relates the Ricci tensors of the two metrics (see, e.g.,
\cite{HAWE}) in terms of ${\cal E}^i_{jk}$.  Using the explicitly
symmetric form of the Ricci tensor, such an identity is:
\begin{equation}
R_{ij}-\bar{R}_{ij} = \bar{D}_k{\cal E}^k{}_{ij}-
\bar{D}_{(i}{\cal E}^k{}_{j)k}+{\cal E}^k{}_{ij}{\cal E}^l{}_{kl}-
{\cal E}^k{}_{l(i}{\cal E}^l{}_{j)k} \,.
\label{riccidec}
\end{equation}

We rewrite the basic evolution system in the following form:
\begin{eqnarray}
\hat{\partial}_t h_{ij} &=& -2NK_{ij} \,, \label{hevol} \\
\hat{\partial}_t K_{ij} &=& -\bar{D}_i \bar{D}_j N + 
 \bar{D}_k (N {\cal E}^k{}_{ij}) 
 - N \bar{D}_{(i}{\cal E}^k{}_{j)k} \nonumber
 \\  &&  + N {\cal E}^k{}_{ij}{\cal E}^l{}_{kl}
 - N {\cal E}^k{}_{l(i}{\cal E}^l{}_{j)k}
 + N (KK_{ij}- 2K_i{}^lK_{jl}) + N  \bar{R}_{ij} \,, \label{kevol}
\end{eqnarray}
where we have used the relation (\ref{dtdb}) to convert covariant
derivatives with respect to the connection $\bar{D}_i$ into covariant
derivatives with respect to $\bar{D}_i$.  Moreover, the Lie
derivatives in the time operator $\hat{\partial}_{t}$ are expanded as
\begin{eqnarray}
{\pounds}_\beta h_{ij} = \beta^k \bar{D}_k h_{ij} +
h_{ik}\bar{D}_j\beta^k+h_{jk} \bar{D}_i\beta^k \,, \nonumber
\end{eqnarray}
and 
\begin{eqnarray}
{\pounds}_\beta K_{ij} = \beta^k \bar{D}_k K_{ij} +
K_{ik}\bar{D}_j\beta^k+K_{jk} \bar{D}_i\beta^k \,. \nonumber
\end{eqnarray}
In the above equations, ${\cal E}^i_{jk}$ is shorthand for
equations~(\ref{eps},\ref{delt}). Hence this is the classic ADM form
of the evolution system, but expressed using a background metric and
connection.

It will be useful here to introduce an alternative expression for
Eq.~(\ref{kevol}). This form of the equation (for Gaussian
coordinates) is given in~\cite{LANDAU} and to write it we introduce
a densitized extrinsic curvature (of weight $1$) 
\begin{equation} 
{\cal K}^{i}{}_{j} = h^{\frac{1}{2}} h^{ik} K_{kj} \,, \label{denextc}
\end{equation}
where $h$ is the determinant of the three-metric $h_{ij}$. The 
evolution of this quantity can be found through the definition of 
the extrinsic curvature (which implies the evolution equations 
$\hat{\partial}_t h = - 2 h N K$ and $\hat{\partial}_t h^{ij} = 
2 N K^{ij}$) 
\begin{eqnarray}
h^{-\frac{1}{2}}h_{il}\hat{\partial}_t {\cal K}^{l}{}_{j} =
- N K K_{ij} + 2 N K_i{}^lK_{jl} + \hat{\partial}_t K_{ij} \,. \nonumber
\end{eqnarray}
Combining this expression with the evolution of $K_{ij}$ [Eq.~(\ref{kevol})]
we get 
\begin{eqnarray}
\hat{\partial}_t {\cal K}^{i}{}_{j}  &=&  
 h^{\frac{1}{2}}h^{ik} \left\{ -\bar{D}_k \bar{D}_j N 
 + \bar{D}_l (N {\cal E}^l{}_{kj}) 
 - N \bar{D}_{(k}{\cal E}^l{}_{j)l}
  + N {\cal E}^l{}_{kj}{\cal E}^m{}_{lm} \right.
 - N {\cal E}^l{}_{m(k}{\cal E}^m{}_{j)l} \left.
 +  N \bar{R}_{k j} \right\}\,, \label{ksimp}
\end{eqnarray}
Equation~(\ref{ksimp}) departs from the standard form of ADM evolution
systems, which explicitly preserve the symmetry of the extrinsic 
curvature and hence only require six components.  Instead here
one obtains the evolution to nine components, which satisfy the three
additional constraints
\begin{eqnarray}
  h_{l[i} {\cal K}^{l}{}_{j]} = 0 \,. \nonumber
\end{eqnarray}
A direct symmetrization, ${\cal K}^{'}_{ij}
= h_{l(i} {\cal K}^{l}{}_{j)}$, e.g., after each evolution step in a
numerical algorithm, would be sufficient for enforcing those constraints.

An {\em evolution} equation for ${\cal E}^i_{jk}$ can 
be computed in the following way: first, using the evolution equation for 
$h_{ij}$ [Eq.~(\ref{hdot})] and the Ricci identities for the commutation of
spatial derivatives
\begin{equation} 
(D_iD_j-D_jD_i){\cal A}^{k\cdots}{}_{l\cdots} =
R^k{}_{mij}{\cal A}^{m\cdots}{}_{l\cdots}+\cdots -
R^m{}_{lij}{\cal A}^{k\cdots}{}_{m\cdots}-\cdots \,, \label{ricc}
\end{equation}
where $R^i{}_{jkl}$ denotes the Riemann tensor of $h_{ij}$, one can find
that
\begin{equation}
\partial_t \Gamma^i{}_{jk} = D_{(j}D_{k)}\beta^i - R^i{}_{(jk)l}\beta^l
+D^i(NK_{jk})-2D_{(j}(NK_{k)}{}^i) \,. \label{patc}
\end{equation}
On the other hand, the Lie derivative of the Christoffel symbols is a
tensor that can be expressed as follows~\cite{SCHO}
\begin{equation}
{\pounds}_\beta \Gamma^i{}_{jk} = D_jD_k\beta^i-R^i{}_{kjl}\beta^l \,.
\label{libc}
\end{equation}
Combining Eqs.~(\ref{patc}) and~(\ref{libc}) we get
\begin{equation}
\hat{\partial}_t \Gamma^i{}_{jk} = D^i(NK_{jk})-2D_{(j}(NK_{k)}{}^i)\,.
\label{cdot}
\end{equation}
which re-expressed for the background geometry reads
\begin{equation}
\hat{\partial}_t {\cal E}^i{}_{jk} = h^{il} \left\{
\bar{D}_{l}(NK_{jk})-2\bar{D}_{(j}(NK_{k)l}) + 2 N {\cal E}^m{}_{jk}
K_{ml}\right\} + \hat{\partial}_t \bar{\Gamma}^i{}_{jk}\,, \label{evee}
\end{equation}
where $\hat{\partial}_t \bar{\Gamma}^i{}_{jk}$ has to be replaced
by the corresponding background version of Eq.~(\ref{cdot}).

With the promotion of the components of the tensor ${\cal E}^i{}_{jk}$ 
to independent variables, we must re-examine the additional constraints 
those variables must satisfy.   The first one concerns the metric 
compatibility of the connection, i.e., $D_{i} h_{jk} =0$ which, expressed 
in terms of the background connection, is equivalent to the definition 
of ${\cal E}^i{}_{jk}$ [Eq.~(\ref{eps})], or to the following relationship
\begin{equation} 
\bar{D}_kh_{ij} = 2{\cal E}_{(ij)k}\,, \hspace{5mm} \mbox{where}
\hspace{3mm} {\cal E}_{ijk} = h_{il}{\cal E}^l{}_{jk}\,.\label{metc}
\end{equation}
Another condition, which can be seen as a consequence of (\ref{metc}),
is the metric-induced property of the Riemann tensor of $h_{ij}$
\begin{eqnarray}
R_{ijkl} + R_{jikl} = 0 \,. \nonumber
\end{eqnarray}
In our framework, this condition expresses the fact that the action of
the commutator of background covariant derivatives, $[\bar{D}_i,\bar{D}_j]$, 
when applied to the physical metric, is given by the usual rule
\[ (\bar{D}_k\bar{D}_l-\bar{D}_l\bar{D}_k)h_{ij} = -2h_{m(i}
\bar{R}^m{}_{j)kl} \,. \]
This in particular contains the relation $\bar{D}_{[i}{\cal E}^k{}_{j]k}
=0$ which has been used to write the differences between Ricci 
tensors~(\ref{riccidec}) in an explicitly symmetric form.

\subsection{Non-linear perturbations}

In the previous subsection we introduced a number of systems which
made explicit use of the existence of a background spatial metric. The
spatial connection tensor ${\cal E}^k{}_{ij}$ arose naturally with the
consideration of a background metric.  In contrast, we will need to
explicitly introduce ``perturbative'' spatial metric and extrinsic 
curvature tensors.  To simplify the notation we will use the
symbol $\mb{\delta}$ to denote ``perturbative'' quantities: 
$\mb{\delta}{\cal A}^{i\ldots}{}_{j\ldots}={\cal A}^{i\ldots}{}_{j\ldots}
-\bar{{\cal A}}^{i\ldots}{}_{j\ldots}$.   Then, the relation between
the ``perturbative'' spatial metric and extrinsic 
curvature tensors
\[ \phi_{ij} = \mb{\delta} h_{ij}=h_{ij}-\bar{h}_{ij}\,, \hspace{4mm}
   \psi_{ij} = \mb{\delta} K_{ij}=K_{ij}-\bar{K}_{ij}\,, \]
is given by 
\begin{eqnarray}
\hat{\partial}_t \phi_{ij}  =  -2N \psi_{ij}\,. \nonumber
\end{eqnarray}
On the other hand, we can always write the following evolution equation
for the background extrinsic curvature 
\begin{equation}
\hat{\partial}_t \bar{K}_{ij} =  -\bar{D}_i\bar{D}_j N + N( \bar{R}_{ij}+
\bar{K}\bar{K}_{ij}-2\bar{K}_i{}^l\bar{K}_{jl}) - \bar{S}_{ij} \,, 
\label{kbdo}
\end{equation}
which has to be considered as the definition of $\bar{S}_{ij}$.
In the special case where ($\bar{h},\bar{K},N,\beta$) is an
exact solution of the Einstein equations, Eq.~(\ref{kbdo}) reduces to
$\bar{S}_{ij}=0$ and can be interpreted as the dynamical 
evolution of an alternative initial data set 
$(\bar{h}^o_{ij},\bar{K}^o_{ij})$.

Subtracting Eq.~(\ref{kbdo}) from Eq.~(\ref{kevol}) eliminates terms
involving second derivatives of the lapse and the background Ricci
curvature and introduces additional extrinsic curvature terms,
\begin{eqnarray}
\hat{\partial}_t \psi_{ij} = \bar{D}_k (N {\cal E}^k{}_{ij}) 
-N\bar{D}_{(i}{\cal E}^k{}_{j)k}+N{\cal E}^k{}_{ij}{\cal E}^l{}_{kl} 
-N{\cal E}^k{}_{l(i}{\cal E}^l{}_{j)k}+N\mb{\delta}(KK_{ij}- 2K_i{}^lK_{jl}) 
+\bar{S}_{ij} \,. \nonumber
\end{eqnarray}
The extrinsic curvature terms can be absorbed with the use of the 
densitized variable ${\cal K}^i{}_j$ [see Eq.~(\ref{denextc})] 
for both the physical and background spacetimes.  Then, we can write
the following alternative equation for the evolution of the components
of the extrinsic curvature
\begin{eqnarray}
\hat{\partial}_t \mb{\delta}{\cal K}^{i}{}_{j} =  
h^{\frac{1}{2}}h^{ik}\left\{ \bar{D}_l(N {\cal E}^l{}_{jk}) 
-N(\bar{D}_{(j}{\cal E}^l{}_{k)l}-{\cal E}^l{}_{jk}{\cal E}^m{}_{lm} 
+{\cal E}^l{}_{m(j}{\cal E}^m{}_{k)l})+
\eta_{kl}\hat{\partial}_t\bar{{\cal K}}^{l}{}_{j}+\bar{S}_{jk}\right\}\,,
\label{knl}
\end{eqnarray}
where 
\begin{eqnarray}
\eta_{ij} = -\mb{\delta}(h^{-\frac{1}{2}} h_{ij})=
(\bar{h}^{-\frac{1}{2}} -  h^{-\frac{1}{2}})\bar{h}_{ij} 
- h^{-\frac{1}{2}} \phi_{ij} \,. \nonumber
\end{eqnarray}
It is worth noting that in equation~(\ref{knl}) background Ricci terms
have been eliminated.  The choice of an exact background solution
would eliminate $\bar{S}_{ij}$ as well.  The term involving
$\hat{\partial}_t \bar{{\cal K}}^{m}{}_{j}$ will be different from
zero in general, but it is homogeneous in the metric deviations from
the background.

The expression of the evolution equation for ${\cal E}^i{}_{jk}$ 
[Eq.~(\ref{evee})] in terms of the ``perturbative'' and background 
quantities is the following
\begin{eqnarray}
\hat{\partial}_t {\cal E}^i{}_{jk} & = & N\left\{ h^{il}(\bar{D}_l\psi_{jk}
-2\bar{D}_{(j}\psi_{k)l})+\mb{\delta}h^{il}(\bar{D}_l\bar{K}_{jk}
-2\bar{D}_{(j}\bar{K}_{k)l}) \right. \nonumber \\
& & \left. +2h^{il}{\cal E}^m{}_{jk}(\psi_{lm}+\bar{K}_{lm})\right\} 
+[h^{il}\psi_{jk}+\phi^{il}\bar{K}_{jk}
-2\psi_{(j}{}^i\delta_{k)}{}^l]\bar{D}_lN \,, \nonumber
\end{eqnarray}
where $\psi_j{}^i=\mb{\delta}K_j{}^i$ and $h^{ij}$ must be understood 
in terms of $\phi_{ij}$ and $\bar{h}_{ij}$, i.e. $h^{ij} = 
([\mb{\phi}+\mb{\bar{h}}]^{-1})^{ij}$.

In summary, utilising the background geometry and identifying the
key geometric variables we have derived four evolution systems,
for the sets of fields $(h_{ij},K_{ij})$, $(h_{ij},{\cal K}_{ij})$, 
$(\phi_{ij},\psi_{ij},{\cal E}^i{}_{jk})$ and
$(\phi_{ij},\mb{\delta}{\cal K}_{ij},{\cal E}^i{}_{jk})$ respectively. 
All those equations are formulated in terms of the background metric 
and connection, but the physical metric also appears explicitly. 
In addition, background source terms, e.g., $\bar{R}_{i j}$, 
$\hat{\partial}_t \bar{\Gamma}^i{}_{jk}$ are present.

\section{3+1 decomposition with background metric: Weak version and
hyperbolicity}\label{sec3}

In what follows we introduce a modified version of the framework
described in Sec.~\ref{sec2}.  As before, in addition to the physical 
metric $h_{ij}$ we assume a second metric field, denoted by $\bar{h}_{ij}$, 
cohabiting with $h_{ij}$ on each of the hypersurfaces $\Sigma(t)$, but
instead of considering the same unit normal vector we consider a new
one $\bar{n}^{\mu}$ which we assume to be parallel to the physical one
$n^\mu$, i.e. $\bar{n}^\mu=\lambda n^\mu$ ($\lambda>0$).  We can combine 
these objects to define a background four-metric in the following way
\begin{eqnarray}
 \bar{g}^{\mu\nu}  =  \bar{h}^{\mu\nu} - \bar{n}^{\mu}\bar{n}^{\nu} =
\bar{h}^{\mu\nu} - \lambda^2 n^\mu n^\nu \,, \nonumber
\end{eqnarray}
where $\bar{h}^{\mu\nu}$ satisfies the same conditions as in the strong
version [see Eq.~(\ref{conhb})].  The condition that $\bar{n}^{\mu}$ is 
a unit normal vector with respect to $\bar{g}_{\mu\nu}=\bar{h}_{\mu\nu}
-\lambda^{-2} n_\mu n_\nu$ ($\bar{h}_{\mu\nu}\bar{n}^\nu=0$) reflects
the fact that the lapse scalar for the background is different from
that of the physical spacetime. We will call it $\bar{N}$ and assume is
a strictly positive function.  Then, we can take $\lambda= N/\bar{N}$
and therefore the relation between normals can be written as
\begin{eqnarray}
\bar{n}^{\mu} = \frac{N}{\bar{N}} n^{\mu} \hspace{4mm} \mbox{and}
\hspace{4mm} \bar{n}_\mu = \frac{\bar{N}}{N} n_\mu \,. \nonumber
\end{eqnarray}
With the assumptions made in this weak version of the 3+1 framework
there is still a unique projection operator onto the hypersurfaces 
$\Sigma(t)$, $h^{\mu}{}_{\nu}=\bar{h}^{\mu}{}_{\nu}= \delta^{\mu}{}_{\nu} 
+n^{\mu}n_{\nu} = \delta^{\mu}{}_{\nu} + \bar{n}^{\mu}\bar{n}_{\nu} $.

Using the relations $\bar{N} = - t^{\mu} \bar{n}_{\mu}$ and
$\beta^{\mu} = t^{\mu} - \bar{N} \bar{n}^{\mu} = t^{\mu} - N n^{\mu}$ 
we write the background line-element as
\begin{eqnarray}
ds^2 = -\bar{N}^2dt^2+\bar{h}_{ij}(dx^i+\beta^idt)(dx^j+\beta^jdt) \,. 
\nonumber
\end{eqnarray}

\begin{figure}
\centerline{\epsfig{figure=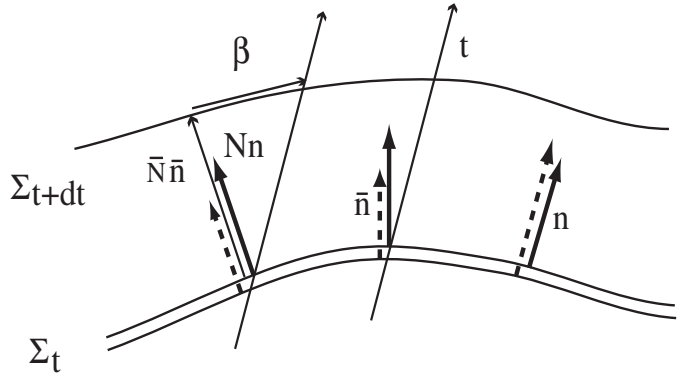,width=3.5in,height=2.0in}}
\caption{3+1 decomposition on a background geometry: weak version.
Two successive level surfaces of the time scalar are shown, with two sets 
of corresponding unit normal vectors (in bold and dashed arrows).  The 
decomposition of the time flow vector field into the lapse and shift 
components is shown for the two possible lapse alternatives. The essence 
of the construction is that in order to advance from the level hypersurface
$\Sigma(t)$ to $\Sigma(t+dt)$ one uses the same time operator 
$\hat{\partial}_{t} = \partial_t - {\pounds}_{\beta}$.}\label{fig2}
\end{figure}

Pictorially, the current geometric construction is presented in 
Fig.~\ref{fig2}.   The Eulerian observers associated with the background 
metric follow the same worldlines as the corresponding observers in the
physical spacetime, but their acceleration is different: 
$\bar{a}_{i}=\bar{D}_{i}(\ln \bar{N})$ as
opposed to $a_{i}=D_{i}(\ln N)$.   Therefore, some of the rigidity in
the correspondence between physical and background spacetimes is lost,
but one still has a unique time evolution operator, $\hat{\partial}_t$,
which is a consequence of having the same shift spatial vector 
$\beta^i$ in both constructions.

As in Sec.~\ref{sec2}, the second fundamental form associated with the 
background geometry can be constructed from the expression (\ref{caex}),
which leads to the usual relation between spatial metric and extrinsic
curvature
\begin{eqnarray}
\bar{K}_{ij}= -\textstyle{1\over2}{\pounds}_{\bar{n}} \bar{h}_{ij}~~
\Rightarrow ~~ \hat{\partial}_t \bar{h}_{ij} = -2\bar{N} \bar{K}_{ij} 
\,. \nonumber
\end{eqnarray}

\subsection{First order systems and hyperbolicity}
The evolution equations for $(h_{ij},K_{ij},{\cal E}^i{}_{jk})$, 
(\ref{hevol}), (\ref{kevol}) and (\ref{evee}), as they
stand do not satisfy any hyperbolicity condition, which implies that we
cannot formulate a well-posed initial-value problem (see~\cite{FRIE,REUL}).
There are many ways of reformulating Einstein's equations in order to get
a hyperbolic system of partial differential equations
(see~\cite{FRIE,ANCY,FRIR} and references therein).  The structure of the
equations for $(h_{ij},K_{ij},{\cal E}^i{}_{jk})$ suggests that the
hyperbolicity properties will be very closely linked to the first-order
symmetric hyperbolic formulation introduced recently in~\cite{ANYO}.
This has been named the ``Einstein-Christoffel'' (EC) system, which
is based in considering combinations of the variables
$(h_{ij},K_{ij},\Gamma^i{}_{jk})$.  In~\cite{KSTC}, implementing this
formulation by means of a pseudospectral collocation scheme, a stable
evolution of spherically symmetric black holes with horizon excision has
been achieved, and in~\cite{KIST}, the total evolution time of 3D black 
hole computations has been extended by using EC type systems.

As in the EC formulation~\cite{ANYO}, instead of considering the lapse
$N$ as a freely specifiable scalar function, we will arbitrarily prescribe
the scalar density (of weight $-1$) $\alpha$, which in the case of
the physical metric is defined through the following expression
\begin{equation}
N(\alpha,h) = \alpha h^{\frac{1}{2}}\,, ~~ \mbox{where}~~
h = \mbox{det}(h_{ij})\,. \label{laps}
\end{equation}
This automatically changes the principal part of the evolution equation
for the extrinsic curvature [Eq.~(\ref{kdot})], which will be crucial
for the hyperbolic structure of the final evolution equations 
(see~\cite{ANCY} for details).
This way of prescribing the gauge precludes the possibility of
choosing the background lapse $\bar{N}$ to be the same as the one of the
physical spacetime $N$, which is the main motivation for the weak 
formulation of the 3+1 framework given above.  Instead, one can introduce 
a background densitized lapse $\bar{\alpha}$ as above
\begin{equation}
\bar{N}(\bar{\alpha},\bar{h}) = \bar{\alpha} \bar{h}^{\frac{1}{2}}\,.
\label{lapb}
\end{equation}

We start the construction of the hyperbolic formulation with the evolution 
equation for the spatial metric ``perturbation'' $\phi_{ij}$, which reads
as
\begin{equation} 
\hat{\partial}_t \phi_{ij} = -2N\psi_{ij}-2\mb{\delta}N\,\bar{K}_{ij}\,,
\hspace{4mm} \mbox{where} \hspace{4mm} \mb{\delta}N=N-\bar{N}\,,
\label{mphv}
\end{equation}
and where $N$ and $\bar{N}$ must be understood as given by the expressions
(\ref{laps}) and (\ref{lapb}) respectively.  Notice that now we have
a source term proportional to the difference of lapses which did not
appear in the previous formulation.

With regard to the evolution equation for the extrinsic curvature
``perturbation'' $\psi_{ij}$, we must first rewrite the term
$\bar{D}_i\bar{D}_jN$ in term of $\alpha$.  Taking into account
the expression for the covariant derivative of a scalar density $\sigma$ 
of weight $w$, $D_i\sigma = \partial_i\sigma-w\Gamma^j{}_{ij}\sigma$,
it becomes
\[ \bar{D}_i\bar{D}_jN = \bar{D}_{(i}(N{\cal E}^k{}_{j)k})+
h^{\frac{1}{2}}\left\{\bar{D}_i\bar{D}_j\alpha+{\cal E}^k{}_{k(i}\bar{D}_{j)}
\alpha\right\}\,.\]
As is clear, this modifies the principal part of the equation
for $K_{ij}$ (\ref{kevol}).  The result can be written as follows
\begin{eqnarray}
\hat{\partial}_t K_{ij} & = & \bar{D}_k(N{\cal E}^k{}_{ij})-
2\bar{D}_{(i}(N{\cal E}^k{}_{j)k})+N({\cal E}^k{}_{ij}{\cal E}^l{}_{kl}+
2{\cal E}^{[k}{}_{k(i}{\cal E}^{l]}{}_{j)l}+KK_{ij}\nonumber \\
& & -2K_i{}^lK_{jl}+
\bar{R}_{ij})-h^{\frac{1}{2}}\bar{D}_i\bar{D}_j\alpha  \,.\label{ephv}
\end{eqnarray}
Then, the equation for $\psi_{ij}$ takes the form
\begin{eqnarray}
\hat{\partial}_t \psi_{ij} & = & \bar{D}_k(N{\cal E}^k{}_{ij})-
2\bar{D}_{(i}(N{\cal E}^k{}_{j)k})+N({\cal E}^k{}_{ij}{\cal E}^l{}_{kl}+
2{\cal E}^{[k}{}_{k(i}{\cal E}^{l]}{}_{j)l})\nonumber \\
& & +\mb{\delta}\left[N(KK_{ij}-2K_i{}^lK_{jl})\right]+\mb{\delta}N\,
\bar{R}_{ij}+\bar{h}^{\frac{1}{2}}\bar{D}_i\bar{D}_j\bar{\alpha}
-h^{\frac{1}{2}}\bar{D}_i\bar{D}_j\alpha \,. \label{evpsi}
\end{eqnarray}
Finally, the equation for ${\cal E}^i{}_{jk}$ is
\begin{eqnarray}
\hat{\partial}_t {\cal E}^i{}_{jk} & = & h^{il}\left\{
\bar{D}_l(NK_{jk})-2\bar{D}_{(j}(NK_{k)l})+2NK_{lm}{\cal E}^m{}_{jk}\right\}
\nonumber \\
& & + \bar{h}^{il}\left\{2\bar{D}_{(j}(\bar{N}\bar{K}_{k)l})
-\bar{D}_l(\bar{N}\bar{K}_{jk})\right\} \,. \label{evne}
\end{eqnarray}
The system of partial differential equations we have got for
$(\phi_{ij}, \psi_{ij},{\cal E}^i{}_{jk})$ [Eqs.~(\ref{mphv}), (\ref{evpsi}),
and (\ref{evne})] is still not explicitly symmetric hyperbolic.  We can get
such form by introducing a new quantity, with the same number of
independent components as ${\cal E}^i{}_{jk}$ (eighteen), which
will be considered as a fundamental variable~\cite{ANYO}
\begin{equation}
\zeta_{kij} = \textstyle{1\over2}\Delta_{kij}-
h_{k(i}h^{lm}(\Delta_{|lm|j)}-\Delta_{j)lm}) =
{\cal E}_{(ij)k}+h_{k(i}h^{lm}({\cal E}_{|lm|j)}-{\cal E}_{j)lm}) \,.
\label{zetd}
\end{equation}
The inverse relations are
\begin{eqnarray}
\Delta_{kij} & = & 2\left\{\zeta_{kij}+2h_{k(i}h^{lm}[\zeta_{j)lm}-
\zeta_{|lm|j)}]\right\} \,, \label{deze} \\
{\cal E}_{ijk} & = & 2\zeta_{(jk)i}-\zeta_{ijk}+4h_{jk}h^{lm}\zeta_{[il]m}
\,. \label{eetz}
\end{eqnarray}

Then, let us look for a system of evolution equations for the unknowns
$(\phi_{ij},\psi_{ij},\zeta_{ijk})$.  As is clear, Eq.~(\ref{mphv})
remains the evolution equation for $\phi_{ij}$.  The evolution
equation for $\psi_{ij}$ is obtained by replacing ${\cal E}^i{}_{jk}$
by $\zeta_{ijk}$ in Eq.~(\ref{ephv}).  The procedure to carry out this 
calculation is to express first  ${\cal E}^i{}_{jk}$ in terms of 
$\Delta_{ijk}$ [Eq.~(\ref{delt})] and then to use the following
important property of $\Delta_{ijk}$
\begin{equation} 
\bar{D}_{[l}\Delta_{k]ij} = -h_{m(i}\bar{R}^m{}_{j)lk} \,, \label{prhe}
\end{equation}
which follows from the Ricci identities~(\ref{ricc}).  The quantity
analogous to $\Delta_{ijk}$ in the EC formulation is 
${\cal G}_{kij} = \partial_kh_{ij}$.  It satisfies the analogous simpler
relation $\partial_{[k}{\cal G}_{l]ij} = 0$.  These relations constitute
one of the keys to find a explicitly symmetric hyperbolic system.
Once the property~(\ref{prhe}) has been used, we have to express 
$\Delta_{ijk}$ in terms of $\zeta_{ijk}$ [Eq.~(\ref{deze})]. After some 
calculations we obtain
\begin{eqnarray}
\hat{\partial}_t \psi_{ij} & = & -N\left\{ h^{kl}\bar{D}_k\zeta_{lij}
-{\cal E}^k{}_{ij}{\cal E}^l{}_{kl}+2{\cal E}^{(k}{}_{k(i}
{\cal E}^{l)}{}_{j)l}
+\textstyle{1\over2}(\bar{R}_{ij}+h^{kl}\bar{R}^m{}_{kl(i}h_{j)m})
\right. \nonumber \\
& & -2h^{kl}h^{mn}\left[ \zeta_{ikm}\zeta_{jln}+
2(\zeta_{kl(i}-\zeta_{(i|kl|})(\zeta_{j)mn}-\zeta_{|mn|j)})
-2\zeta_{(i|km|}(\zeta_{j)ln}-\zeta_{|ln|j)})
 \right. \nonumber \\
& & \left.\left. +2\zeta_{kij}(\zeta_{lmn}-\zeta_{mnl}) \right]\right\}
+\mb{\delta}\left[N(KK_{ij}-2K_i{}^lK_{jl})\right]+\mb{\delta}N\,\bar{R}_{ij}
\nonumber \\
& & +h^{\frac{1}{2}}({\cal E}^k{}_{ij}\bar{D}_k\alpha
-2{\cal E}^k{}_{k(i}\bar{D}_{j)}\alpha-\bar{D}_i\bar{D}_j\alpha)
+\bar{h}^{\frac{1}{2}}\bar{D}_i\bar{D}_j\bar{\alpha} \,,
\label{sdot}
\end{eqnarray}
where ${\cal E}^i{}_{jk}$ must be understood as given by
expression~(\ref{eetz}).

To find the evolution equation for $\zeta_{kij}$ we have to apply
$\hat{\partial}_t$ to the expression~(\ref{zetd}) and use all the
information we have.  It turns out that in order to have the desired
form for the principal of this equation one needs to make use of the
momentum constraint~(\ref{mcon}) that the physical spacetime has to satisfy. 
After some manipulations the result is
\begin{eqnarray}
\hat{\partial}_t\zeta_{kij} & = & -\bar{D}_k(NK_{ij})+4NK_{k(i}h^{lm}
[\zeta_{j)lm}-\zeta_{|lm|j)}]+2Nh_{k(i}[K_{j)}{}^lh^{mn}-\delta_{j)}{}^l
K^{mn}]{\cal E}_{lmn} \nonumber \\
& & +4h^{\frac{1}{2}}h_{k(i}\delta_{j)}{}^mh^{ln}K_{l[m}[\bar{D}_{n]}\alpha+
\alpha{\cal E}^p{}_{n]p}]+h_{k(i}\bar{D}_{j)}(\bar{N}\bar{K}) +
h_{l(i}[\bar{D}_{j)}(\bar{N}\bar{K}_k{}^l)\nonumber \\
& & + \bar{D}_{|k|}(\bar{N}\bar{K}_{j)}{}^l)-
\bar{D}^l(\bar{N}\bar{K}_{j)k})]- h_{k(i}h_{j)n}h^{lm}
[2\bar{D}_l(\bar{N}\bar{K}_m{}^n)-
\bar{D}^n(\bar{N}\bar{K}_{lm})]\,. \label{zdot}
\end{eqnarray}
At this point we have a system of first-order evolution
equations for the unknowns $(\phi_{ij},\psi_{ij},\zeta_{kij})$
[Eqs.~(\ref{mphv}), (\ref{sdot}) and (\ref{zdot})], where  
some quantities constructed from $(h_{ij},K_{ij})$ appear explicitly
but they must be considered as given by their expressions in terms
of $(\phi_{ij},\psi_{ij})$ and $(\bar{h}_{ij},\bar{K}_{ij})$. 
We finish this section by showing the differential structure of
the system (\ref{mphv},\ref{sdot},\ref{zdot}).  This analysis
is standard (see, e.g., \cite{FRIE}) and one only needs to consider
the principal part of the equations, which for our system looks as 
follows
\begin{eqnarray}
\hat{\partial}_t\phi_{ij} & \approx & 0 \,, \nonumber \\
\hat{\partial}_t\psi_{ij} & \approx & -Nh^{kl}\bar{D}_k\zeta_{lij}\,,
\nonumber \\
\hat{\partial}_t\zeta_{kij}  & \approx & -N\bar{D}_k\psi_{ij} \,,
\nonumber
\end{eqnarray}
where $\approx$ means equality up to terms that do not contain
derivatives of the unknowns $(\phi_{ij},\psi_{ij},\zeta_{kij})$.
From the structure of the principal part we deduce that the system is
symmetric hyperbolic.  Indeed, the equations that determine the
kernel of the principal symbol associated with the system are
\begin{eqnarray}
& &(\mb{n}\cdot\mb{\xi})\phi_{ij} = 0 \,, \nonumber \\
& &(\mb{n}\cdot\mb{\xi})\psi_{ij}+\xi^k\zeta_{kij} = 0 \,, \nonumber \\
& &(\mb{n}\cdot\mb{\xi})\zeta_{kij}+\xi_k\psi_{ij}  = 0 \,,\nonumber
\end{eqnarray}
where $\mb{\xi}=(\xi_t,\xi_i)$ is an arbitrary covector, 
$\mb{n}\cdot\mb{\xi}=g^{\mu\nu}n_\mu\xi_\nu=n^\mu\xi_\mu$, and 
$\xi^i=h^{ij}\xi_j$.  From the structure 
of the principal symbol it follows that the characteristic directions of
propagation of the system are the normal to the spacelike hypersurfaces,
$\mb{n}\cdot\mb{\xi}=0$, and the light cone determined by the physical
metric $(\mb{n}\cdot\mb{\xi})^2=1$, and therefore,  as in the EC
formulation, the characteristic speeds are the speed of light and the 
zero speed.  From the analysis of the kernel of the principal symbol we
deduce that the fields propagating along the normal to the hypersurfaces
$\Sigma(t)$ are all the components of $\phi_{ij}$ and 12 components of
$\zeta_{kij}$, namely
\[ \zeta_{kij} = A_k\xi_i\xi_j+2B_{k(i}\xi_{j)}+C_{kij} \,, \]
where $A_i$, $B_{ij}$, and $C_{kij}$ satisfy: $\xi^iA_i = \xi^iB_{ij}
= \xi^iC_{kij}=0$.  The fields propagating along the physical null
cone are all the components of $\psi_{ij}$ and the 6 components
of $\zeta_{kij}$ given by
\begin{eqnarray}
\psi_{ij} & = & D\xi_i\xi_j+2E_{(i}\xi_{j)} + F_{ij} \,, \nonumber \\
\zeta_{kij} & = & \epsilon (D\xi_k\xi_i\xi_j+\xi_kF_{ij}+
2\xi_k\xi_{(i}E_{j)}) \,, \nonumber
\end{eqnarray}
where now $\epsilon=\mb{n}\cdot\mb{\xi}$ ($\epsilon = \pm 1$) and
the quantities $E_i$ and $F_{ij}$ are such that $\xi^iE_i=
\xi^iF_{ij}=0$.

Finally, as it is usual in the theory of symmetric hyperbolic PDEs, we 
can introduce an energy norm for our system [Eqs.~(\ref{mphv}), 
(\ref{sdot}) and (\ref{zdot})] which is given by 
\[ \mbox{E}(t) = \int_{\Sigma(t)} (\phi^{ij}\phi_{ij}+\psi^{ij}\psi_{ij}+
\zeta^{kij}\zeta_{kij})\,\mb{d\Sigma} \,, \]
where $\mb{d\Sigma}$ is the volume element of $\Sigma(t)$ and indices
are raised using the inverse spatial physical metric $h^{ij}$.

\section{Summary and discussion}\label{rand}

\begin{figure}
\centerline{\epsfig{figure=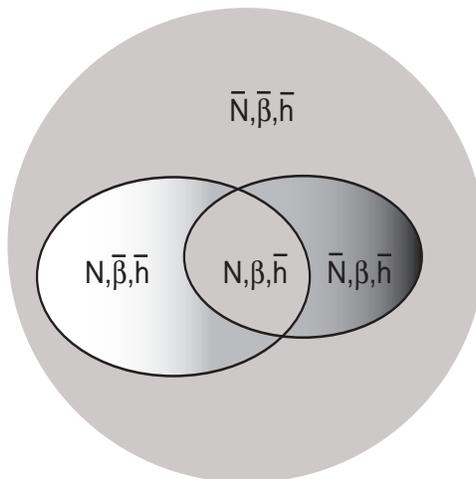,width=2.5in,height=2.5in}}
\caption{The logical relationship between different possible
arrangements for the background geometry. In the most general case
(enclosing set), the background geometry has both its gauge functions
(lapse and shift) different from the physical geometry. The use of a
common lapse {\em or} shift vector defines the two inner sets. For the
overlap of those two sets, both lapse and shift are the same as for
the physical spacetime. This arrangement corresponds to the framework
in Section II (strong version). The developments in Section III (weak 
version) refer to the set on the right, characterised by 
$(\bar{N},\beta,\bar{h})$.}\label{fig3}
\end{figure}

The main results of this paper are the introduction of a set of new
evolution systems for the Einstein equations which explicitly separate
a background geometry. We have studied two broad classes, which differ
slightly in their geometrical construction.  An important realisation
has been that for constructing a hyperbolic set of evolution equations
one must relax the assumption of a common lapse. The reason lies with
the fact that the use of a densitized lapse is important for inducing
the appropriate modification to the principal part of the
equations. This introduces a gauge function that incorporates
information about the three metric.  On the other hand, having a
background spacetime which shares the same lapse and shift as the
physical one is crucial for a maximal simplification of the 
resulting equations. The choices are summarised in
Fig.~\ref{fig3}.  The introduction of a background geometry makes all 
the systems explicitly three-covariant.  Besides a certain 
aesthetic appeal, this feature may be
handy in applications involving multiple coordinate systems, e.g.,
spherical coordinate patches adapted to black holes and infinity
respectively.
       
We note that in determining the light-cone structure, or in the case
of the first order systems, the propagation of eigenfields, it is the
physical metric that is always the governing field. This implies that
there would be no unpleasant side-effects associated with the choice
of a background spacetime that is causally incompatible with the
physical one.  We have not elaborated on the Hamiltonian and
momentum constraints. Those are essential for the construction of
consistent initial data.  The conformal techniques and decompositions
involved in reducing those equations to well behaved elliptic systems
seem at first glance not to be particularly benefiting from the
assumption of a generic background geometry.

A key issue that must be considered is the way in which the use of a 
background geometry would modify the numerical properties of evolution 
equations for the Einstein system. Common to most approaches to the 
evolution problem is the solution of a quasi-linear system of explicitly 
or implicitly hyperbolic equations {\em with sources}. For spacetimes
involving compact objects, large curvature gradients are
present. Those gradients survive in the stationary (non-dynamical
limit) and involve a balance between flux terms in the principal part
of the equation and the source terms. The source terms are not
homogeneous in the evolution fields which can also be expressed as the
existence of longitudinal modes. For fluid spacetimes (e.g., neutron
stars) this balance has physical underpinnings in the fluid pressure
supporting the star. For black holes it is geometric in nature.

The implementation of the systems developed here can proceed in a
number of manners. After a suitable choice of background geometry, a
direct discretization of the equations using standard discrete
approximations is possible. A practise usually adopted is to ignore
the constraint equations in the construction of the algorithm (free
evolution) and to monitor them as time dependent error norms. The
integration of a physical data set will then consist of a sequence of
discrete timesteps into the future. In principle, for a stable and
convergent algorithm, the solution can be found at any future finite
time to any desired accuracy given enough resolution.  With the use of
prescribed background geometry one has the flexibility of a more
interactive approach to the time evolution, namely through an {\em
iteration} of the evolution process. This technique could be useful in
circumstances in which the background geometry is constructed
analytically on the basis of a few poorly known parameters, e.g.,
post-Newtonian locations and momenta of black holes. The iterations
would then provide improved estimates for those parameters. Whereas
{\em any} of the iterative steps can in principle converge to the true
solution, it has been our argument that the residual errors will be
dramatically reduced as one improves the guess on the background
spacetime. 

The incorporation of matter dynamics has not been discussed
here. There is no specific obstruction to the process. In contrary, an
implementation of closely related non-linear perturbation concepts to
the relativistic Euler equations is already underway with promising
results~\cite{SPA}.  An optimal application candidate would be the study 
of non-linear perturbations of a single black hole spacetime. In this
case the stationary exact black hole solutions provide readily
available first guesses for the background geometry.  More
speculative, but well motivated, is the application to multiple black
hole systems. Here the appropriate construction of an approximate
spacetime would make heavy use of post-Newtonian solutions. The
framework provides a well defined path for systematically {\em
incorporating} post-Newtonian results into the full PDE evolution,
which only hinges on the suitable expansion of post-Newtonian results 
into global metric data.

\[ \]
{\bf Acknowledgements:} We warmly thank M. Bruni, P. Laguna, and 
R. Maartens for comments on the manuscript.  CFS acknowledges 
support from the European Commission (contract HPMF-CT-1999-00149).  
PP acknowledges support from the EU Programme ``Improving the Human 
Research and Socio-Economic Knowledge Base'' (Research training network 
HPRN-CT-2000-00137) and the Nuffield Foundation (award NAL/00405/G).


\end{document}